\documentclass[showpacs,floatfix,aps,prl,twocolumn,superscriptaddress,amssymb,nobibnotes ]{revtex4-2}

\usepackage{amsmath}
\usepackage{amssymb}
\usepackage{amsfonts}
\usepackage{latexsym}
\usepackage{graphicx}
\usepackage{color}
\usepackage{cleveref}
\usepackage{microtype}

\usepackage{glossaries}
\setacronymstyle{long-short}
\newacronym{hBN}{hBN}{hexagonal boron nitride}
\newacronym{AFM}{AFM}{atomic force microscope}
\newacronym{PFM}{PFM}{piezo force microscopy}
\newacronym{2D}{2D}{two-dimensional}
\newacronym{LFM}{LFM}{lateral force microscopy}
\newacronym{bMLG}{bMLG}{bent monolayer graphene}
\newacronym{bBLG}{bBLG}{bent bilayer graphene}

\glsresetall 

\begin{document}

\title{Programming moiré patterns in 2D materials by bending}

\author{Mäelle~Kapfer}
\thanks{These authors contributed equally to this work.}
\affiliation{Department of Physics, Columbia University, New York, NY, USA}

\author{Bjarke~S.~Jessen}
\thanks{These authors contributed equally to this work.}
\affiliation{Department of Physics, Columbia University, New York, NY, USA}

\author{\\Megan~E.~Eisele}
\affiliation{Department of Physics, Columbia University, New York, NY, USA}

\author{Matthew~Fu}
\affiliation{Department of Physics, Columbia University, New York, NY, USA}

\author{Dorte~R.~Danielsen}
\affiliation{Center for Nanostructured Graphene, Technical University of Denmark, DK-2800, Denmark}
\affiliation{DTU Physics, Technical University of Denmark, DK-2800, Denmark}

\author{Thomas~P.~Darlington}
\affiliation{Department of Mechanical Engineering, Columbia University, New York, NY, USA}

\author{Samuel~L.~Moore}
\affiliation{Department of Physics, Columbia University, New York, NY, USA}

\author{Nathan~R.~Finney}
\affiliation{Department of Mechanical Engineering, Columbia University, New York, NY, USA}

\author{Ariane~Marchese}
\affiliation{Department of Mechanical Engineering, Columbia University, New York, NY, USA}

\author{Valerie~Hsieh}
\affiliation{Department of Physics, Columbia University, New York, NY, USA}

\author{Paulina~Majchrzak}
\affiliation{Department of Physics and Astronomy, Aarhus University, 8000 Aarhus C, Denmark}

\author{Zhihao~Jiang}
\affiliation{Department of Physics and Astronomy, Aarhus University, 8000 Aarhus C, Denmark}

\author{Deepnarayan~Biswas}
\affiliation{Department of Physics and Astronomy, Aarhus University, 8000 Aarhus C, Denmark}

\author{Pavel~Dudin}
\affiliation{Synchrotron SOLEIL, Université Paris-Saclay, F-91192 Gif sur Yvette, France}

\author{José~Avila}
\affiliation{Synchrotron SOLEIL, Université Paris-Saclay, F-91192 Gif sur Yvette, France}

\author{Kenji~Watanabe}
\affiliation{National Institute for Materials Science, 1-1 Namiki, Tsukuba, 305-0044, Japan}

\author{Takashi~Taniguchi}
\affiliation{National Institute for Materials Science, 1-1 Namiki, Tsukuba, 305-0044, Japan}

\author{Søren~Ulstrup}
\affiliation{Department of Physics and Astronomy, Aarhus University, 8000 Aarhus C, Denmark}

\author{Peter~B{\o}ggild}
\affiliation{Center for Nanostructured Graphene, Technical University of Denmark, DK-2800, Denmark}
\affiliation{DTU Physics, Technical University of Denmark, DK-2800, Denmark}

\author{P.~J.~Schuck}
\affiliation{Department of Mechanical Engineering, Columbia University, New York, NY, USA}

\author{Dmitri~N.~Basov}
\affiliation{Department of Physics, Columbia University, New York, NY, USA}

\author{James~Hone}
\affiliation{Department of Mechanical Engineering, Columbia University, New York, NY, USA}

\author{Cory~R.~Dean}
\affiliation{Department of Physics, Columbia University, New York, NY, USA}

\date{\today}

\maketitle

\textbf{Moiré superlattices in twisted two-dimensional materials have generated tremendous excitement as a platform for achieving quantum properties on demand. However, the moiré pattern is highly sensitive to the interlayer atomic registry, and current assembly techniques suffer from imprecise control of the average twist angle, spatial inhomogeneity in the local twist angle, and distortions due to random strain. Here, we demonstrate a new way to manipulate the moiré patterns in hetero- and homo-bilayers through in-plane bending of monolayer ribbons, using the tip of an atomic force microscope. This technique achieves continuous variation of twist angles with improved twist-angle homogeneity and reduced random strain, resulting in moiré patterns with highly tunable wavelength and ultra-low disorder. Our results pave the way for detailed studies of ultra-low disorder moiré systems and the realization of precise strain-engineered devices.}

\glsresetall 

\begin{figure*}[tb]
	\includegraphics[width=1\linewidth]{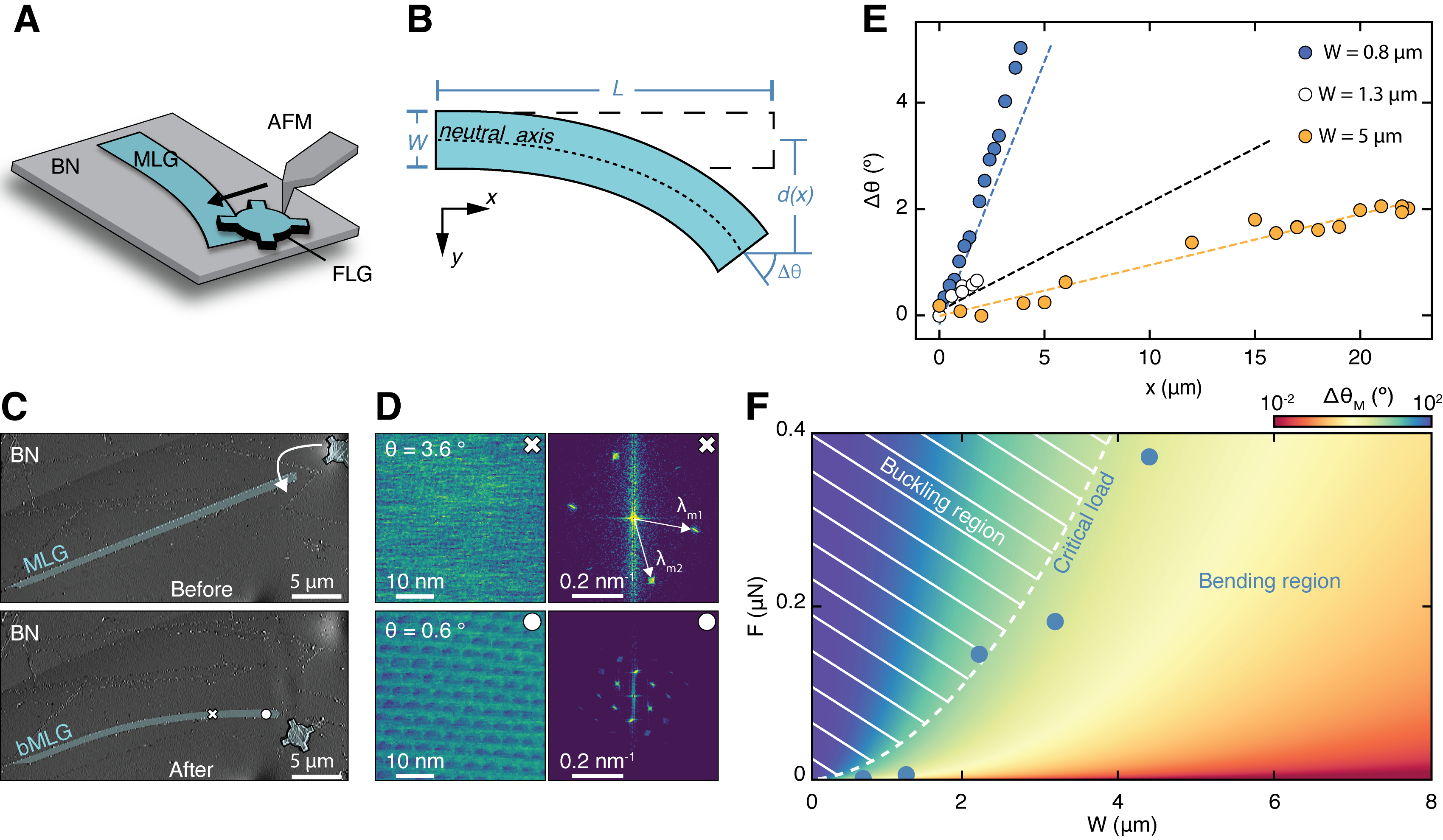}
	\caption[]{\textbf{Mechanical bending of \gls{2D} materials and twist-angle control:} 
	A) Schematic of bending of a \gls{2D} material ribbon using a nano-manipulator and the tip of an \gls{AFM}. 
    B) Schematic of a \gls{2D} material ribbon as a beam of length $L$ and width $W$, indicating a neutral axis. A coordinate system, $(x,y)$, marks the initial position of the ribbon. Deflections from the unbent neutral axis are denoted by $d(x)$, resulting in a twist angle $\Delta\theta$ from the initial position.
 	C) \gls{AFM} images of a \gls{bMLG} on \gls{hBN} before (top) and after (bottom) bending. 
	D) \gls{PFM} scans (left) taken at the marked spots indicated in C) and their corresponding FFTs (right). 
 	E) Extracted twist angle as a function of the $x$ position defined in B) for different widths of \gls{bMLG} on \gls {hBN}, highlighting the continuously changing twist angle. The dashed lines are the calculated twist angles for a classical cantilever bending model with no fitting parameters.
 	F) Calculation of the maximum relative twist angle from classical cantilever bending theory as a function of the width of the ribbon, $W$, and the load applied at the free end, $F$. Markers correspond to measured buckle points and the dashed line of constant maximum strain of $\pm$ 2.5\,\% for \gls{bMLG} on \gls{hBN}. 
}\label{fig:1}
\end{figure*}
\begin{figure*}[tb]
	\includegraphics[width=0.9\linewidth]{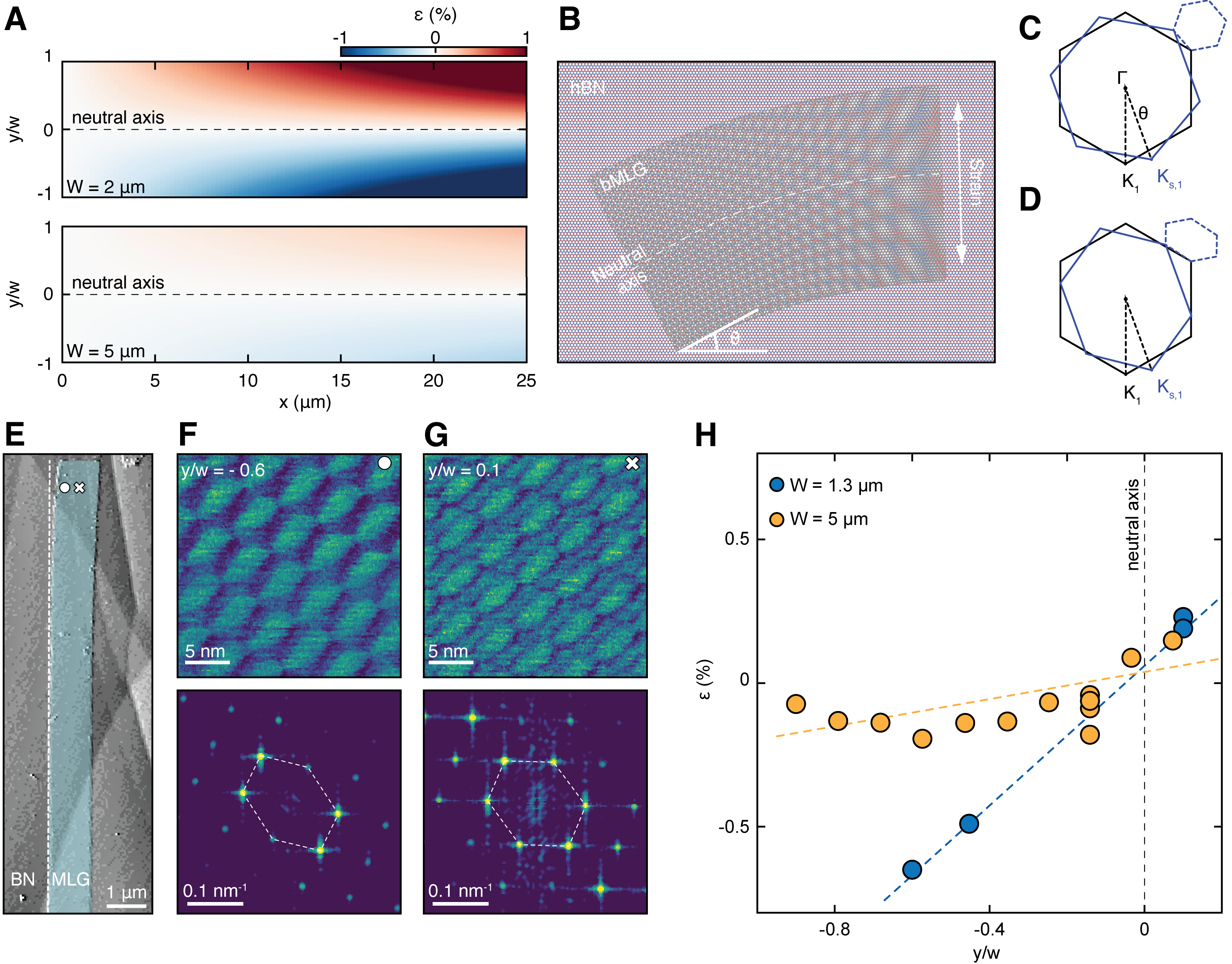}
	\caption[]{\textbf{Tunable strain gradients:} 
	A) Calculation of the strain distribution in a 25\,$\mu$m long, 2\,$\mu$m (top) or 5\,$\mu$m (bottom) wide \gls{bMLG} deflected by $d(L)=$ 300\,nm. The dashed line shows the unstrained neutral axis. 
	B) Sketch of a distorted moiré superlattice, highlighting the effect of strain in the low-angle limit. For moderate twist angles (left side, $\theta = 27^{\circ}$), the strain has only a minute effect on the fidelity of the moiré superlattice, while for low twist-angles (right side, $\theta = 2^{\circ}$), the lattice is heavily distorted for both compressive and tensile strain.
	C,D) Brillioun zones (BZ) of two twisted graphene layers at no strain and tensile strain, respectively, highlighting the sensitivity of the superlattice BZ towards strain.
 	E) \gls{AFM} image of a thin \gls{bMLG} on \gls{hBN}. The dashed line indicates the position prior to bending.
 	F,G) \gls{PFM} scans taken across the \gls{bMLG} in panel E), with positions of the \gls{PFM} scans marked on the \gls{AFM} image of the sample. The lower panels show the corresponding FFTs, highlighting the increasingly distorted moiré pattern away from the neutral axis of the sample.
 	H) Summary of the strain gradient across a thin (blue) and wide (yellow) \gls{bMLG} as a function of the normalized $y$ position. The dashed lines are the strain gradient predicted by beam bending theory for the corresponding geometry and deflection.
}\label{fig:2}
\end{figure*}

\begin{figure*}[tb]
	\includegraphics[width=0.9\linewidth]{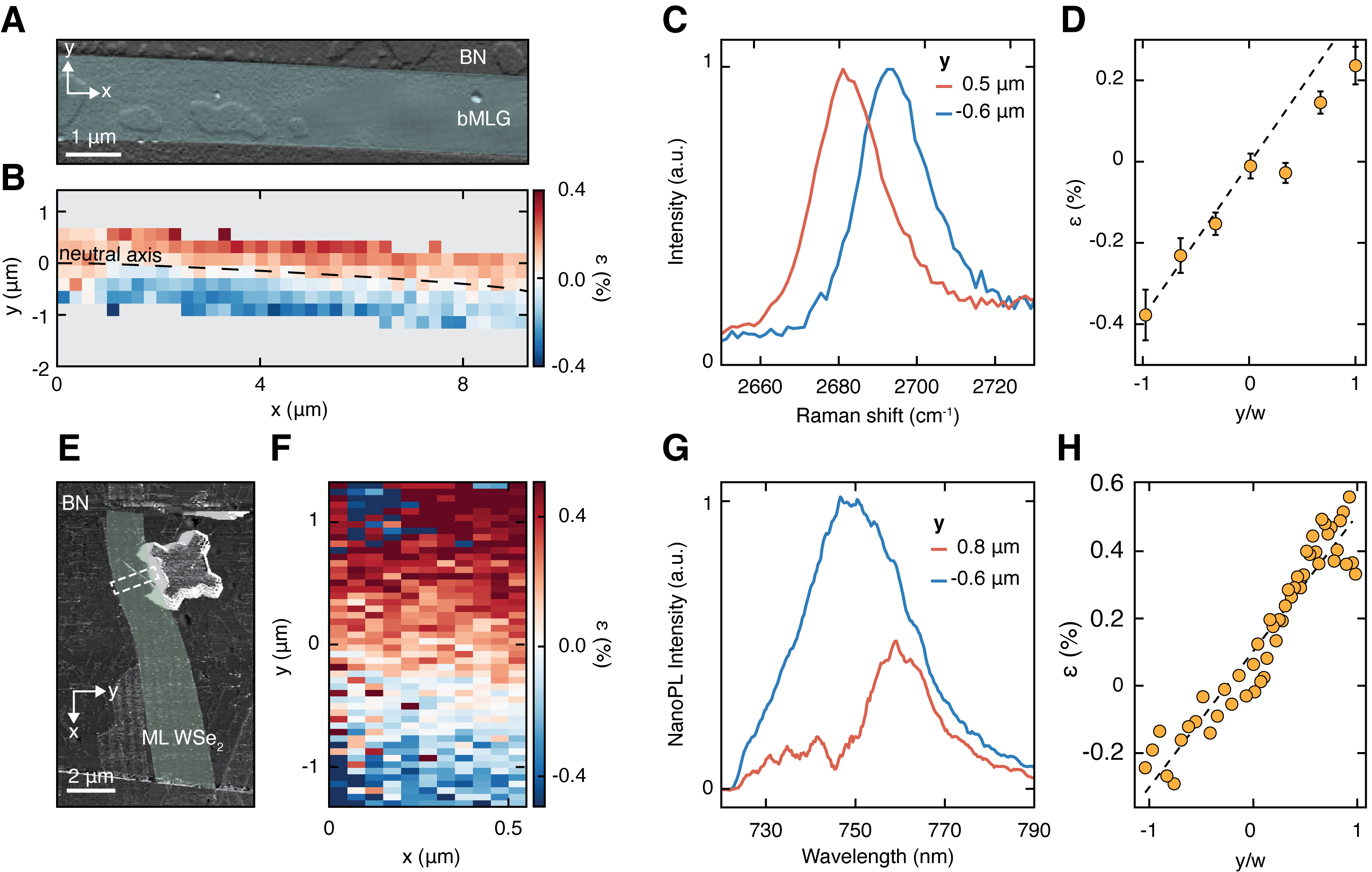}
	\caption[]{\textbf{Characterization of strain without moiré magnification:} 
	A,B) The upper panel shows an \gls{AFM} image of a 1.6\,$\mu$m wide \gls{bMLG}, while the lower panel shows the corresponding strain map as extracted from Raman spectroscopy. Red indicates tensile strain while blue indicates compressive strain, with a white neutral axis identified in the middle of the \gls{bMLG}.
	C) Raman spectra at fixed twist angle but varying strain for the \gls{bMLG} shown in A). 
	D) Extracted strain of the sample in A), with the dashed line being the beam bending model for a 1.6\,$\mu$m wide beam deflected by $d(L)$= 300\,nm yielding a strain gradient of 0.64\,\% per $\mu$m. 
	E,F) The left panel shows an \gls{AFM} image of a 2.4\,$\mu$m wide ribbon of monolayer WSe$_2$ on \gls{hBN}. The right panel shows a strain map of the sample, from the dashed area in E). The strain is extracted from nano-PL mapping, where the peak position is related to changes in band alignment due to strain.
	G) Representative spectra of areas of compressive (blue) and tensile (red) strain in the WSe$_2$.
	H) Extracted strain vs. distance from the neutral axis of the sample in E), with the dashed line being a linear fit yielding a strain gradient of 0.33\,\% per $\mu$m.
	}\label{fig:3}
\end{figure*}

\begin{figure*}[tb]
	\includegraphics[width=0.9\linewidth]{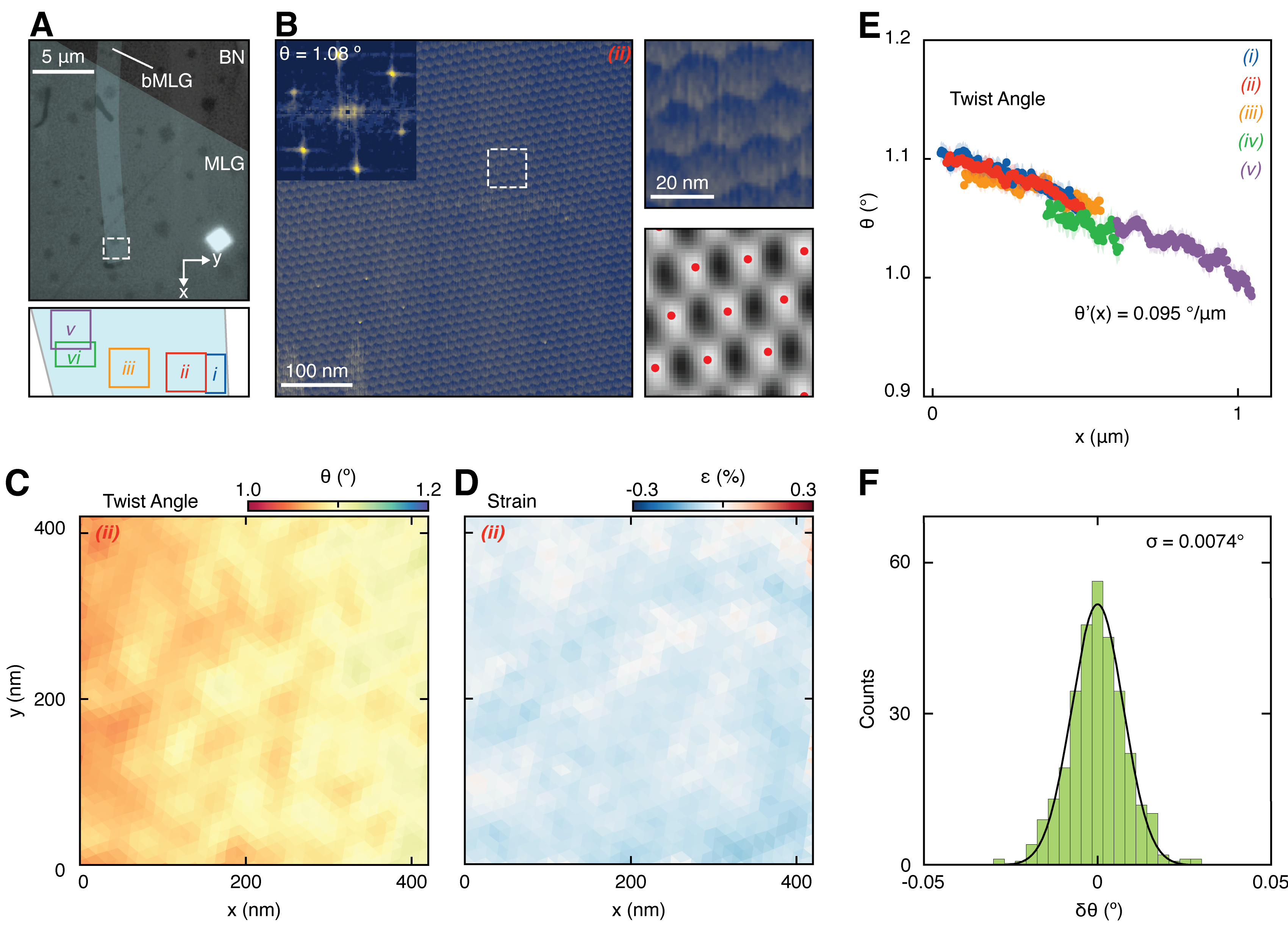}
	\caption[]{\textbf{Ultra-low disorder twisted bilayer graphene:} 
	A) Optical microscope image of a \gls{bBLG} sample. The dashed square indicating the position of detailed \gls{LFM} scans, and further highlighted in the lower panel. 
	B) Large area \gls{LFM} scan of the \gls{bBLG} sample, with the corresponding FFT in the inset. A global twist angle of 1.08$^{\circ}$ is extracted from the FFT peak positions. A series of such scans along the $x$-direction of the ribbon reveals a uniform global twist-angle gradient of 0.095$^{\circ}$ per $\mu$m (see SI). The top right panel is a 53\,x\,53\,nm$^2$ zoom-in of the \gls{LFM} scan. The bottom right panel is a non-local means denoised result of the zoom-in, where the red dots indicate the center of the moiré sites used for local twist-angle and strain extraction. 
	C) Map of the local twist angle calculated from the moiré sites positions of \gls{LFM} scan area (\textit{ii}). A twist-angle gradient is visible along the $x$-direction of the map, with an average twist angle of 1.08$^{\circ}$. 
	D) Map of the local strain of \gls{LFM} scan area (\textit{ii}). The average strain of the map is at a low value of around $0.05\,\%$.
	E) Local twist-angle evolution as a function of the $x$-position for \gls{LFM} scan areas (\textit{i}) to (\textit{v}), revealing a local twist-angle gradient of 0.095$^{\circ}$ per $\mu$m, identical to the extracted global twist-angle gradient.
	F) With an established twist-angle gradient, we extract the residual local twist-angle for all five scans in A), yielding an intrinsic twist-angle disorder of 0.0074$^{\circ}$.
	}\label{fig:4}
\end{figure*}

\glsresetall 

The interference pattern between twisted or lattice-mismatched layers of \gls{2D} materials leads to a periodic electrostatic scattering potential superimposed over the atomic lattice \cite{li_observation_2010,yankowitz_emergence_2012}, as well as periodic structural distortions due to in-plane lattice relaxation and out-of-plane deformations \cite{woods_commensurate-incommensurate_2014,li_imaging_2021,yankowitz_dynamic_2018}. The most dramatic consequence of this so-called moiré pattern is the appearance of flat bands in the electronic energy spectrum, which are found to host a variety of interacting-driven many-body states such as correlated insulators, superconductivity, and magnetic order as well as non-trivial topology \cite{balents_supercon_2020,mak2022semiconductor}. The ability to laminate 2D materials with different atomic composition and lattice constants, and at arbitrary twist angle, allows for wide tunability of the moiré wavelength, amplitude and symmetry \cite{noauthor_moire_nodate}, with further modification possible through in-situ rotation \cite{ribeiro-palau_twistable_2018,hu_-situ_2022,inbar_quantum_2022}, pressure \cite{yankowitz_dynamic_2018,yankowitz_tuning_2019}, and strain \cite{qiao_twisted_2018,bi_designing_2019,gao_heterostrain_2021}. Moiré patterning in \gls{2D} materials therefore provides a promising and flexible new platform to realize correlated physics and topology in quantum materials by design \cite{noauthor_moire_nodate}.

Small variations in the moiré pattern, however, can lead to dramatically different behaviours, making it difficult to uncover the exact relationship between structural details of the superlattice and resulting electronic properties. For example, the flat band hosting superconductivity in twisted bilayer graphene emerges for twist angles between the layers within only $\pm0.1^{\circ}$ of $1.1^{\circ}$ \cite{bistritzer_moire_2011-1,cao_nematicity_2021}, corresponding to a difference in moiré wavelength of just $\pm$1\,nm. This places stringent demands on the fabrication process \cite{kim2016van}. Moreover experimental studies have found that spatially varying twist angle and distortion of the moiré unit cell due to random strain fluctuations are ubiquitous when using current assembly techniques \cite{uri_mapping_2020,grover_chern_2022,choi_correlation-driven_2021,turkel_orderly_2022}.
Such non-uniformity compromises the ability to deterministically engineer specific band structures through well defined moiré patterns or even correlate experimental measurement with predictive theoretical modeling \cite{wagner_global_2022,pantaleon_tunable_2021,de2022imaging,nakatsuji2022moire}. The inability to control this ``moiré disorder'' and the lack of any systematic understanding of its influence represents a critical roadblock toward continued progress in this field \cite{lau2022reproducibility}.

Here, we demonstrate a new technique that addresses these issues by using global strain fields to manipulate local moiré structures. We use an \gls{AFM} to bend monolayers of 2D materials in-plane, resulting in a continuously varying twist angle along the length of the ribbon. We confirm through real-space imaging using \gls{PFM} and \gls{LFM}, as well as optical mapping via Raman spectroscopy and nano-photoluminescence (nanoPL), that the resulting twist-angle and strain gradients vary smoothly and with a predictable spatial dependence that matches a simple beam-bending model. The ability to vary the total twist angle by several degrees allows us to span wide-ranging moiré length scales in a single device. Most surprising, we find dramatically reduced moiré disorder in the bent ribbon geometry with, for example, magic-angle twisted bilayer graphene showing a twist-angle variation of less than 0.0074$^{\circ}$ and strain fluctuation less than 0.023\,\% over $1\,\mu$m$\,\times\,2.5\,\mu$m. Our results suggest that manipulation of the global strain field offers a new route toward controlling moiré disorder in \gls{2D} heterostructures.

Figure 1A shows a schematic of the bending process, illustrated for the case of a \gls{bMLG} ribbon on an \gls{hBN} substrate. The heterostructure is first assembled using the dry transfer technique where we use the \gls{hBN} to sequentially pick up a straight graphene ribbon, followed by a few-layer graphite manipulator shaped into a gear-like geometry (see methods for details). Using an \gls{AFM} tip to slide the manipulator over the end of the ribbon allows us to apply an in-plane load \cite{ribeiro-palau_twistable_2018}. In response, the ribbon bends at the load-end, with the far end effectively clamped due to interfacial friction.

We characterize the deformed ribbon using an $(x,y)$ coordinate system, where $x$ follows the neutral axis prior to bending, and $y$ is the distance from the neutral axis perpendicular to $x$, as shown in Fig. 1B. We also introduce $W$ and $L$ as the width and length of the ribbon, respectively, $d(x)$ as the deflection from the initial neutral position, and $\Delta\theta(x)$ as the local twist angle relative to the orientation of the unbent ribbon. For small deflections, $\Delta\theta(x)$ relates to $d(x)$ and $L$ according to 
\begin{equation}
\Delta\theta(x) \approx \tan^{-1}(d(x)/L). \label{eq:0}
\end{equation}
An \gls{AFM} image of a \gls{bMLG} on \gls{hBN} is shown in Fig. 1C, with the \gls{bMLG} highlighted in blue, showing the initially unbent (top) and final bent (bottom) geometries. The bending process is both reversible and robust as we can bend and unbend the ribbon up to several degrees without evidence of plastic deformation (see SI). Upon releasing the load, static friction is sufficient to secure the deformation profile of the bent beam, and we can even continue to use the dry transfer technique to add additional layers to the heterostructure without negatively affecting the geometry. We have applied the bending geometry to varying combinations of graphene, \gls{hBN}, and transition metal dichalcogenides and observed similar results in all cases. 

Due to the closely matched lattice constants of graphene and \gls{hBN}, a large wavelength moiré superlattice emerges in the small twist angle regime, with wavelength, $\lambda$, varying with the twist angle, $\theta$, according to 
\begin{equation}
\lambda= \frac{{\left( {1 + \delta } \right)a}}{{\sqrt {2\left( {1 + \delta } \right)\left( {1 - \cos \left( \theta \right)} \right) + {\delta ^2}} }}, \label{eq:1}
\end{equation}
where $a$ is the graphene lattice constant, $\theta$ is the twist angle, and $\delta$ is the lattice mismatch (0.017 for graphene on \gls{hBN}) \cite{yankowitz_emergence_2012}. Thus, measurement of the moiré wavelength provides a way to determine the local twist angle \cite{mcgilly_visualization_2020,zheng_robust_2016}. Figure 1D shows real-space and Fourier-space images acquired by PFM from two different locations on the \gls{bMLG} shown in Fig. 1C. From the positions of the peaks in the FFT, we extract the average moiré wavelength and use Eq. \ref{eq:1} to determine the average twist angle within the scan window. Repeating this measurement along multiple points on the \gls{bMLG}, we spatially map the twist angle along the ribbon. Figure 1E shows the measured change in twist angle, $\Delta\theta$, as a function of $x$ for three different ribbons widths of 0.8$\,\mu$m, 1.3$\,\mu$m, and 5$\,\mu$m. In each case, the twist angle varies approximately linearly with position, with a slope that inversely correlates with ribbon width.

To gain insight into the interplay between ribbon geometry and resulting twist-angle variation, we model the system as a cantilever with one end clamped and a point-load applied to the free end. While either end of our ribbons is, in principle, free to move, we find in practice that interfacial friction keeps most of the ribbon static, effectively serving as a clamp. For this textbook problem, the deflection along the cantilever is given by 
${d\left( x \right) = \frac{{F{x^2}}}{{6EI}}\left( {3L - x} \right)}$,
where $F$, $E$, and ${I=tW^3/12}$ are the point-load, Young's modulus, and geometrical moment of inertia with beam thickness $t$, respectively. Dashed lines in Fig. 1E are calculated from this model together with Eq. \ref{eq:0}, assuming previously measured mechanical properties of graphene \cite{lee_measurement_2008} and with no other free parameters. While cantilever models can explicitly include friction \cite{stupkiewicz1994elastic,mogilevsky1997plane}, we find excellent agreement with our experimental data when inverting the $x$-direction, confirming the validity of this simpler model. Based on this understanding, we show in Fig. 1F a calculated map of the maximum achievable twist angle as a function of load and ribbon width for experimentally relevant values. Although the map in Fig. 1F shows angles ranging up to 90$^{\circ}$ twist, in practice, we observe that beyond a critical load, the \gls{bMLG} buckles \cite{lit_bending_2015}, suddenly transitioning to two straight sections separated by a highly strained fold (see SI). 

The experimentally observed buckling load for different ribbon widths are indicated by blue circles in Fig. 1F. By balancing the van der Waals energy between the ribbon and the substrate with the elastic energy from strain, we calculate a critical strain of $\sim$ 2.5\,\%, after which buckling becomes energetically favorable, shown as the dashed line in Fig. 1F (see SI for details). Despite not taking moiré effects into account, this model predicts the buckling points well and sets an upper limit on $\Delta\theta$ of approximately 5$^{\circ}$ for a 3\,$\mu$m wide \gls{bMLG}.

Assuming again a classical beam-bending model, we expect a strain gradient across the \gls{bMLG}, given by ${\varepsilon = -y/\rho(x)}$, where ${\rho(x) = {\left( {1 + {{\left( {d\left( x \right)'} \right)}^2}} \right)^{3/2}}/d\left( x \right)''}$ is the local radius of curvature. Figure 2A shows the calculated spatial strain maps for two 25\,$\mu$m long ribbons, one $2\,\mu$m wide and one $5\,\mu$m wide, each with a maximum displacement of $d(L)=$ 300\,nm. As expected, the \gls{bMLG} is in a state of compressive strain on the inside radius and tensile strain on the outside radius, with the neutral axis separating the two regimes. In both cases, the strain evolves linearly across the \gls{bMLG} with a coefficient that also grows linearly towards the end of the \gls{bMLG}. The strain gradient reaches as high as 1.7\,\% per $\mu$m for the $2\,\mu$m wide ribbon but only 0.1\,\% per $\mu$m for the $5\,\mu$m wide ribbon. This indicates that the bending geometry allows independent tuning of twist angle and strain gradient through choice of ribbon width; a wide ribbon can be used to achieve low strain values, even when varying the total angle by several degrees. On the other limit, deflecting a narrow ribbon can yield large local strain values while spanning overall similar twist angles. We confirm this independent tuning through the use of (nano) angle-resolved photoemmision spectroscopy, where we spatially map out the band structure of a 4\,$\mu$m wide \gls{bMLG} bent 4$^{\circ}$. As expected from a low-strain ribbon, we observe a significant BZ rotation but otherwise minimal impact on the band structure (see SI).

Analysis of the moiré pattern allows us to map the local strain field, in addition to twist angle, since the moiré pattern magnifies any distortion in the constituent lattices \cite{C4FD00146J}. This is visualized in Fig. 2B, where we show a schematic of a bent \gls{bMLG} on \gls{hBN}. 
The impact of the strain is apparent from the heavily distorted moiré lattice \cite{Miller_structural_2010,halbertal_moire_2021,hou_evaluation_2022}, especially visible for the large moiré periods in the low twist-angle regime. The effect is maximal at the top and bottom edges of the \gls{bMLG} and gradually decreases as we move towards the unstrained neutral axis, near the middle. Figure. 2C and 2D show the resulting Brillouin zones (BZs) of the twisted layer in the unstrained and tensile regions, respectively.

Figure 2E-G shows examples of distorted moiré patterns acquired by \gls{PFM} from a 1.3\,$\mu$m wide \gls{bMLG} on \gls{hBN}. The dashed line in Fig. 2E shows the initial position of the \gls{bMLG} (highlighted in blue) before bending. Figures 2F,G show high-resolution moiré patterns (upper panels), and corresponding FFTs (lower panels), acquired from two locations (indicated by white symbols) of the bBLG shown in Fig. 2E. The scan windows correspond to two different $y$ positions (i.e., different strain points), but fixed $x$ positions (i.e., same twist angle). Following Refs. \onlinecite{kerelsky_maximized_2019, halbertal_extracting_2022}, the average of the reciprocal lattice vectors measured from the FFT gives the local twist angle, whereas the ratio of the reciprocal vectors gives a measure of the local strain.
The pattern in Fig. 2F, acquired close to the neutral axis, gives $\theta$ $\sim$ 1.37$^{\circ}$ and $\varepsilon$ $\sim$ 0.2\,\%, while Fig. 2G, taken in the compressive part of the \gls{bMLG}, gives $\theta$ $\sim$ 1.34$^{\circ}$ and $\varepsilon$ $\sim$ -0.65\,\%. 

Figure 2H shows a plot of the strain vs. normalized $y$ position $y/w$, where $y$ is the distance from the neutral axis and $w=W/2$, the half-width of the \gls{bMLG} (blue circles). The strain gradient is approximately linear, with a slope that matches the value of $\sim$1\,\% per $\mu$m calculated from our model (dashed blue line) with no free parameters. For comparison, strain vs. $y/w$ from a wider 5\,$\mu$m ribbon is also shown (yellow circles). Again we find good agreement between the strain field measured from the distortions in the moiré pattern and our model (dashed yellow line). Moreover, as expected, the wider ribbon gives a substantially reduced strain gradient, confirming the ability to use the \gls{bMLG} width to tune the strain independently of the twist angle. 

So far, we have focused on \gls{bMLG}s with a visible moiré pattern. However, for large twist angles (>5$^{\circ}$) the resulting moiré wavelength is below our resolution limit. For those systems the strain can instead be probed from optical measurements. Figure 3A shows a \gls{bMLG} on \gls{hBN} where no moiré was detected in the \gls{PFM} scans, indicating a twist angle larger than 10$^{\circ}$ \cite{mcgilly_visualization_2020}. The corresponding strain map extracted from Raman spectroscopy is seen in Fig. 3B. The strain is calculated from the position of the 2D peak, where we expect a shift of $\sim$27\,cm$^{-1}$ per \% of strain \cite{ni_uniaxial_2008}. Spectra taken along various positions to the neutral axis are shown in Fig. 3C, with the red and blue curves corresponding to tensile and compressive strain, respectively. Using multiple peak positions at a constant $x$-position, we extract the strain evolution along $y$ (Fig. 3D) with the dashed line showing the strain gradient expected for a 1.6\,$\mu$m wide beam deflected by $d(L)=$ 300\,nm, again showing good agreement with a simple cantilever model.

Figure 3E shows a bent monolayer WSe$_2$ ribbon on \gls{hBN}, demonstrating the capability to apply this technique to \gls{2D} materials other than graphene. Because of the significant lattice mismatch between these two materials, no moiré superlattice is observed. We instead use nanoPL to determine the strain in the WSe$_2$ (see SI for details). When subject to strain, the direct band gap of WSe$_2$ is expected to shift by -50\,meV per \% \cite{desai_strain-induced_2014}. Figure 3F is a spatial map of the strain distribution in the WSe$_2$ ribbon calculated from the shift of the peak wavelength (Fig. 3G). 
Similar to \gls{bMLG}, we observe a strain gradient, shown in Fig. 3H, evolving from tensile (blue) to compressive (red) strain and with a linear slope matches theory (Fig. 2H)

Finally, we exploit the bent geometry to fabricate twisted bilayer graphene with continuously varying twist angles through the magic angle. An optical image of the \gls{bBLG} is shown in Fig. 4A (see SI for details of the device fabrication). We perform \gls{LFM} scans along the ribbon and report the results for five different scan areas, all near the region where the magic angle is reached. The scan locations, shown schematically at the bottom of Fig. 4A, span in total 1\,$\mu$m in the $x$-direction and 2.5\,$\mu$m in the $y$-direction. In Fig. 4B, we show one of these scans with the corresponding FFT in the inset, while the remaining can be found in SI. The real-space image shows a remarkably uniform moiré pattern with little evidence of varying twist angle or notable distortion. Similar high-quality moiré patterns were observed in each of the scan regions. We extract the moiré wavelength from the FFTs and use Eq. \ref{eq:1} to calculate the twist angle. Repeating this for different positions along the ribbon, we find that the twist angle between the layers varies from $\theta$ = 2.5$^{\circ}$ in the unbent region to $\theta$ = 1$^{\circ}$ at the end of the \gls{bBLG} (see SI) with an approximately linear gradient of 0.095$^{\circ}$ per $\mu$m.

While we can extract an average twist angle over a single \gls{LFM} scan from its FFT, the striking high quality of our scans opens the possibility to analyze the real-space moiré pattern and extract the local twist angle and strain by locating the center of each moiré sites (lower right panel of Fig. 4B, see SI). The spatial distributions of twist angles and strain, extracted from Fig. 4B, are shown in Fig. 4C and Fig. 4D, respectively. While the overall twist-angle variation remains small ($\pm$ 0.05$^{\circ}$), we see a gradient of the twist angle along the $x$-axis (bending axis) and minimal evolution along the $y$-axis. For the strain, we observe highly uniform values of around -0.05\,\% over the whole area, in line with the visually uniform pattern seen in Fig. 4B. 

In order to characterize the apparent twist-angle gradient, we average twist angle values along the $y$-axis and report this average as a function of the position along the $x$-axis, or bending axis. We repeat this averaging for all five scans and report the twist-angle evolution in Fig. 4E. We observe a uniform gradient of the twist angle of 0.095$^{\circ}$ per $\mu$m -- identical to the global twist-angle gradient found with FFT peak positions. The identical local and global twist-angle gradients confirm that the twist-angle evolution along the \gls{bBLG} is smooth and continuous over the whole \gls{bMLG}. 
Having established the bend-induced twist-angle gradient as the primary source of angle variation in our samples, we set out to estimate the intrinsic twist-angle disorder as a deviation from the global twist-angle gradient. The resulting distribution of the gradient-corrected twist angles is presented in Fig. 4F. We find an intrinsic disorder value of 0.0074$^{\circ}$, which is around three times lower than results obtained with conventional twisted bilayer graphene samples prepared using tear and stack method \cite{uri_mapping_2020,kerelsky_maximized_2019}.

The ability to precisely tune the twist angle and strain within a \gls{2D} heterostructure, in the absence of uncontrolled distortions, paves the way for moiré band structure engineering in the disorder-free limit, including the exciting possibility that moiré patterning can be used as a generalized quantum simulation platform to study strongly correlated physics and topology in quantum materials \cite{noauthor_moire_nodate}. We note that the dramatically reduced moiré disorder observed in our bent geometry is not yet understood. We conjecture that this may relate to the lattice relaxation dynamics in the presence of an externally applied strain field, but further theoretical and experimental work will be required to fully understand both the origin of this behavior and how this interplay may be exploited to realize new control opportunities. Finally, we note that the reversible in-plane bending geometry that we demonstrate, realized through local mechanical actuation, provides a new approach towards generalized strain engineering \cite{amorim_novel_2016,rold_strain_2015} beyond moiré patterning.

\section{Methods}
See Supplemental Information.

\section{Acknowledgments}
The authors acknowledge Jason Li at Oxford Instrument Asylum Research for experimental assistance with the LFM imaging, and acknowledge Dorri Halbertal and Simon Turkel for helpful discussions. 

Fabrication and characterization of the bent ribbon structures was primarily supported as part of Programmable Quantum Materials, an Energy Frontier Research Center funded by the U.S. Department of Energy (DOE), Office of Science, Basic Energy Sciences (BES), under award DE-SC0019443. Fabrication of bent TMD structures was supported by the Columbia MRSEC on Precision-Assembled Quantum Materials (PAQM) - DMR-2011738. C.R.D.and J.H. acknowledge additional support from the Gordon and Betty Moore Foundation’s EPiQS Initiative, grant GBMF10277. D.N.B. is a Moore Investigator in Quantum Materials EPIQS GBMF9455 and the Vannevar Bush Faculty Fellow ONR-VB: N00014-19-1-2630. P.B. and D.R.D. acknowledge support from Danish National Research Foundation (DNRF) Center for Nanostructured Graphene (DNRF103) and EU Graphene Flagship Core 3 (881603). 
The nanoARPES experiments were carried out at the ANTARES beamline of Synchrotron SOLEIL under proposal number 20210204, and supported by Independent Research Fund Denmark under the Sapere Aude program (Grant No. 9064-00057B) and VILLUM FONDEN under the Young Investigator Program (Grant No. 15375).

\section{Author Contributions}
M.K., B.S.J., M.E.E., and C.R.D. wrote the manuscript with input from all authors.
M.K., B.S.J., and N.R.F. prepared the graphene devices, performed AFM, PFM, and LFM measurements, and M.K. and B.S.J. analyzed the data.
T.P.D. performed Raman measurements and T.P.D., B.S.J., and M.K. analyzed the data.
M.E.E. prepared the TMD device. M.F. and S.L.M. performed nanoPL measurements and together with M.E.E. analyzed the data.
D.R.D. calculated the critical buckling strain.
M.K., M.F., D.R.D. and A.K.M. did the beam bending calculations.
B.S.J. and M.K. prepared samples for nanoARPES, P.M., Z.J., S.U., B.S.J., M.K., D.B., J.A., and P.D. performed the measurements, while P.M., Z.J., and S.U. analyzed the data.
V.H. assisted with LFM measurements.
K.W. and T.T. grew and provided the hexagonal boron nitride crystals.
\section{Competing Interests}
The authors declare no competing interests. 

\bibliographystyle{naturemag_noURL}

\onecolumngrid
\pagebreak
\newpage
\widetext
\begin{center}
\textbf{\large Supporting Information: Programming moiré patterns in 2D materials by bending}
\end{center}
\setcounter{equation}{0}
\setcounter{figure}{0}
\setcounter{table}{0}
\setcounter{page}{1}
\makeatletter
\renewcommand{\theequation}{S\arabic{equation}}
\renewcommand{\thefigure}{S\arabic{figure}}
\renewcommand{\bibnumfmt}[1]{[S#1]}
\renewcommand{\citenumfont}[1]{S#1}


\section{Methods}
\subsection{Nano-manipulator fabrication}

Nano-manipulators are made using a bulk-like (>20\,nm thickness) van der Waals material, usually graphite, mechanically exfoliated onto SiO$_2$/Si chips. We find nano-manipulators made from graphite to be more mechanically roboust compared to e.g. hBN. Conventional electron-beam lithography (EBL) and lift-off is used to deposit a metal layer of Cr/Pd (2/50\,nm). We find that the use of Pd is critical to achieve reliable and repeatable nano-manipulation, as softer metals such as Au tend to deform when pushed with the AFM. Subsequently we use the deposited metal as an etch mask to remove all remaining material on the chip, leaving is with metal-coated nano-manipulators. 

\begin{figure*}[h]
	\includegraphics[width = 1\textwidth]{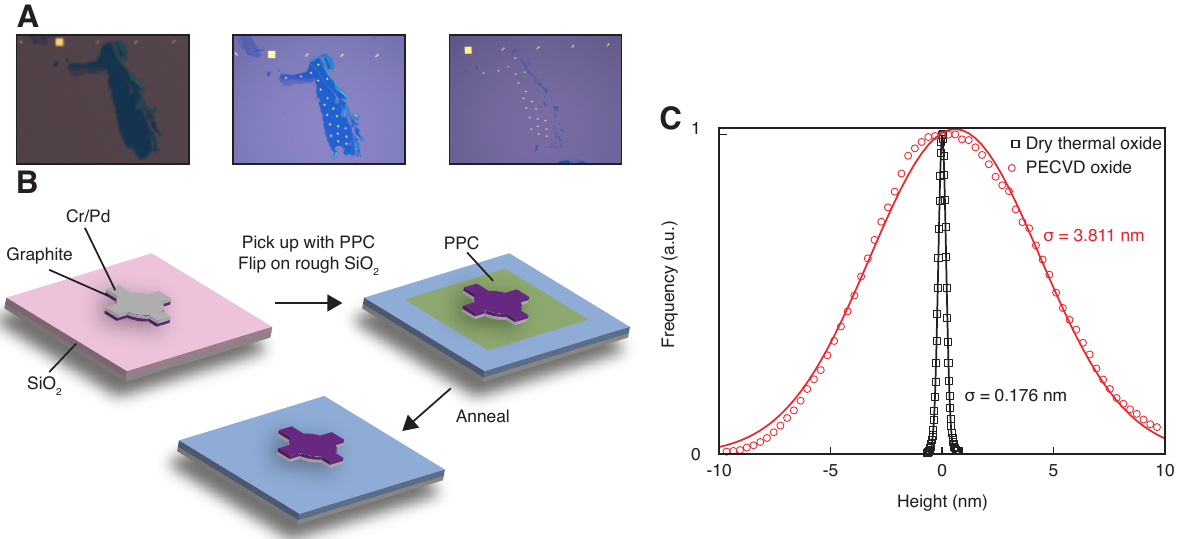}
	\caption[]{\textbf{Nano-manipulator fabrication}
	A) Microscope images of the nano-manipulators fabrication. The left panel shown exfoliated graphene in SiO$_2$, after EBL, Cr/Pd is deposited on the graphite flake (middle panel). After lift-off, the metal is used as a mask to etch the uncovered graphite only leaving behind metal coated manipulators (right panel)
	B) Schematic of the process to flip the nano-manipulator onto rough SiO$_2$, easing further pick-up steps. The nano-manipulator is picked up from the initial SiO$_2$ using a polymer stamp of PPC. The PPC is flipped onto a substrate of rough SiO$_2$, and subsequently thermally decomposed in either high vacuum at >275\,ºC or in an reducing atmosphere (Ar/H$_2$) at >350\,ºC.
	C) Histogram of roughness of PECVD grown SiO$_2$ compared to commercial dry thermal SiO$_2$, as measured by AFM over a 500 x 500\, $\mu$m$^2$ area.}\label{fig:SI_nanoman_fab}
\end{figure*}

Next, the nano-manipulators are picked up with a conventional micro-manipulator using a glass-slide containing a stack of PPC / transparent tape / PDMS. Here the transparent tape serves to keep any contamination from the PDMS away from the PPC. After the nano-manipulator is picked up, the PPC is gently pealed off from the glass slide, and placed with the nano-manipulator facing up on a separate high-roughness SiO$_2$/Si (see below). The PPC is then annealed away, either in a UHV chamber at temperatures above 275\,$^{\circ}$C, or in an Ar/H2 forming gas atmosphere at temperatures above 350\,$^{\circ}$C, effectively flipping the nano-manipulator \cite{polshyn_quantitative_2018, zeng_high-quality_2019}. The nano-manipulator has now been “flipped”, with the vdW material side facing up, and can be placed on our target sample using conventional dry stacking methods.

\subsection{Rough SiO$_2$ fabrication}
We find that the yield of picking up the nano-manipulators drastically increase if they are placed on a rough substrate. To this end, we fabricate substrates with plasma-enhanced chemical vapour deposition (PECVD) grown SiO$_2$. These substrates have a root mean squared roughness of 3.8\,nm, compared to ~0.18\,nm for the usually used thermally grown SiO$_2$. These samples are grown in an Oxford PlasmaPro NGP80 at the following growth conditions: 300\,$^{\circ}$C, 50\,W, 1500\,mTorr. The growth time is 7 minutes, resulting in approximately 300\,nm SiO$_2$.

\subsection{Bilayer process flow}
\begin{figure*}[h]
	\includegraphics[width=1\linewidth]{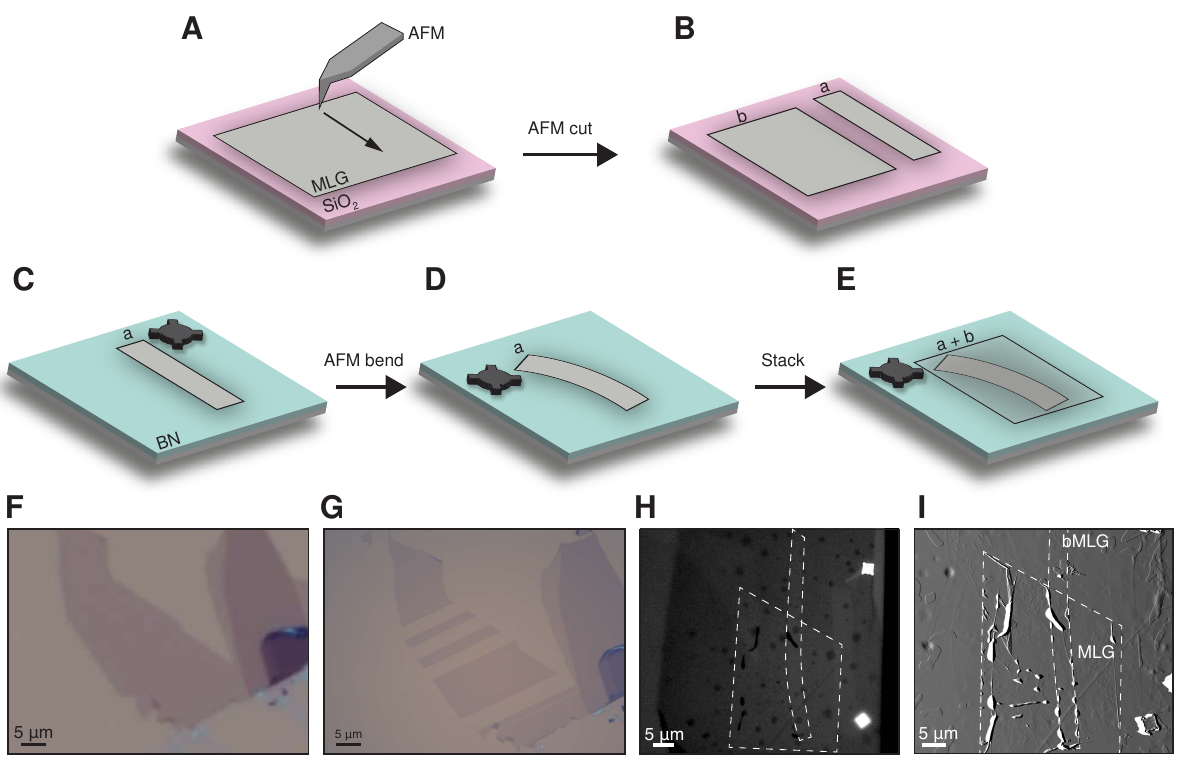}
	\caption[]{\textbf{bBLG fabrication}
	A-E) Schematics of bBLG cutting and fabrication.
	A) AFM shaping of monolayer graphene.
	B) After shaping two flakes remain, a ribbon shaped \textit{a} and a rectangular one \textit{b}.
	C) The ribbon-shaped flake is picked up with hBN and a graphite nano-manipulator is subsequently picked up.
	D) The ribbon is bent using an AFM.
	E) After bending the flake \textit{b} is piked up resulting in a continuously varying twist angle between \textit{a} and \textit{b}.
	F) MLG flake on 90\,nm SiO$_2$.
	G) Same MLG flake as F) after AFM cutting.
	H) Microscope image of the BLG device, the bMLG and rectangular MLG pieces are highlighted with white dashed lines.
	I) AFM image of the same devices shown in H).}\label{fig:SI_tBLGfab}
\end{figure*}

In the following section we describe the fabrication details of making samples of bent bilayer graphene. We start with mechanical exfoliation of graphene onto SiO$_2$/Si chips and identify monolayers from their optical contrast \cite{jessen2018quantitative}. Suitable flakes are then cut using an AFM \cite{li2018electrode} into two separate pieces; one flake and one ribbon (flake a and b in Fig. \ref{fig:SI_tBLGfab}, respectively). We leave a gap of at least 5\,$\mu$m between the two pieces to avoid accidental pick-up during stacking. Using standard dry-stacking techniques \cite{wang2013one, pizzocchero_hot_2016, purdie_cleaning_2018}, we first pick-up the ribbon with hBN and then a nano-manipulator. The nano-manipulator is then used to gently twist the ribbon and moved to the side. Finally, we pick up the remaining larger flake and align it to the untwisted part of the ribbon. At this stage we employ PFM or LFM measurements to determine the local twist-angle and strain profiles, as described in the main text. Next, the sample can either be further encapsulated or flipped onto SiO$_2$ for further scanning probe techniques. Somewhat surprisingly, we always find that the ribbon stays in its twisted state, even further encapsulation and processing.

\subsection{Piezo force microscopy}
PFM measurements were performed on a Bruker Dimension Icon with a Nanoscope V Controller using single frequency excitation at the resonance peak ($\sim$300\,kHz for vertical and $\sim$800\,kHz for lateral). The AC bias voltage was kept below 1\,V. 

\subsection{Lateral force microscopy}
LFM measurements were performed on a Jupiter Asylum AFM in Oxford Instruments Asylum Research facility in Santa Barbara with the help of Jason Li. The samples were scanned at 90$^{\circ}$ and the lateral channel was recorded. The lever used was made out of Si coated with Al, with a spring constant of $\sim$2\,N/m and a radius of curvature of 7\,nm. Typically the setpoint was kept below 0.5\,V resulting in normal forces lower than 10\,nN. 

\subsection{Raman}
Raman measurements were performed on a Nikon Eclipse microscope with a 100x, 0.95 NA air objective. The sample was excited using a linear polarized HeNe laser (Melles Griot) with an excitation wavelength of 633\,nm and a power of 800\,$\mu$W. The sample was raster scanned with a PI piezo scanner with an integration time of 8 seconds per pixel. Back-scattered Raman spectra were collected with the same objective and filtered with 1x notch and 1x Raman edge filters (Semrock) before being passed to the confocal pinhole of 100\,$\mu$m in diameter. Spectra were imaged using a Teledyne PI SRS 300 spectrometer with a 500\,nm blazed, 1200\,gr/mm grating, imaged onto and EMCCD (Teledyne ProEM). 
\subsection{Nano-photoluminescence}
Nano-photoluminescence imaging is performed with a Neaspec Gmbh near-field microscope, with the signal are collected in forward scattering. For optimal tip enhancement, scans are operated with Ag-coated Next Tip EASY-TERS tips in contact-mode.  At each tip position, a spectrum is collected in contact, followed by a spectrum with the sample lowered 150\,nm.  A parabolic mirror (NA = 0.4) collimates the tip-enhanced emission, whereby the light is filtered with a 715\,nm Semrock long pass filter and a 950\,nm Semrock shortpass filter to remove excitation (700\,nm) and AFM deflection (980\,nm) laser light.  Before entering the Andor Kymera spectrometer, the emission is focused through a 75\,$\mu$m pinhole to reduce background emission. The light is then focused onto a 800\,nm, 150\,l/mm grating and refocused onto a Newton CCD, where the spectrum is recorded.

\subsection{NanoARPES}

NanoARPES measurements were carried out on a wide bMLG on hBN device. As shown in the main text in Figure 3, for such large ribbon we expect a small strain gradient. This small strain should not induce any change in the graphene band structure allowing to study only the change in twist angle.

The nanoARPES measurements were performed at the ANTARES beamline of Synchrotron SOLEIL, France. The sample was annealed in ultrahigh vacuum at 343\,K for nine hours prior to the exposure to the beam. During the experiment, the sample was kept at the temperature of 65\,K. \\
A photon energy of 95\,eV was selected and the synchrotron beam was focused with a Fresnel zone plate to a spot size of 1\,$\mu$m. The energy and angular resolution of the MB Scientific A-1 analyser were set to $\sim$40\,meV and 0.1$^{\circ}$, respectively. The $(E, k_x, k_y)$-dependent photoemission spectra were collected using the deflector mode of the MBS analyser, and used to align the analyser to the $K$-point of graphene in the region away from the nano-rotator. The four-dimensional $(E, k, x, y)$-dependent photoemission intensity maps were obtained by raster scanning the sample position with piezoelectric stages and measuring $(E, k)$-snapshot at each position. \\
The local twist angle distribution, $\theta(x, y)$, of bent graphene was extracted by fitting Lorentzian line profiles to the energy distribution curves (EDCs) at fixed momentum through the graphene Dirac cone for different sample positions. A simulation of graphene $E(k)$-dispersion based on a tight-binding model was applied to provide a calibration between the bending angle and EDC peak positions.

\section{Critical buckling strain calculation}
For 2D monolayers out-of-plane buckling is more energetically favorable than in-plane buckling \cite{LEE2016595}, and we therefore focus on out-of-plane buckling (hereafter referred to as buckling). To model buckling of graphene ribbons, we combine two analytical models of buckling of graphene nanoribbons on gold substrates \cite{korhonen2016limits} and buckling of vdW layers \cite{Koskinen_2013}. When the graphene ribbon buckles in-plane elastic energy, $\Delta {E_{el - in}}$, is released at the cost of out-of-plane elastic energy, $\Delta {E_{el - out}}$, and vdW energy cost, $\Delta {E_{vdW}}$, of separating the two layers at the fold \cite{korhonen2016limits,Koskinen_2013}. For the in-plane and out-of-plane elastic energies, we use the following expressions, derived in Ref. \cite{korhonen2016limits}: 
\[\Delta {E_{el - in}} = \frac{{wk\varepsilon {\pi ^2}{A^2}}}{{4\lambda }}, \quad \Delta {E_{el - out}} = \frac{{w{A^2}D{\pi ^4}}}{{4{\lambda ^3}}}.\]
Here, $w$ is the width of the graphene ribbon, $k=19\,$eV/Å$^2$ is the 2D elastic modulus for graphene and $D=1.0$\,eV is the elastic bending modulus for graphene, $A$ is the height of the buckling fold, and $\lambda$ it the width of the buckling fold \cite{korhonen2016limits}. The buckling fold height profile is modeled by $y\left( l \right) = A{\sin ^2}\left( {l\pi /\lambda } \right)$. Before buckling, we assume that the in-plane strain $\varepsilon$ is constant across the inner and outer half of the ribbon, respectively \cite{korhonen2016limits}. Though this is a crude approximation for micron-wide ribbons it still provides a useful starting point. 
 
Using the Hamaker model \cite{HAMAKER19371058}, the vdW interaction energy per area, $U$, can be found by integrating over all pairwise atomic interactions between the graphene and hBN layers, which are modeled by the carbon-boron and carbon-nitrogen Lennard-Jones potentials \cite{NorioInui2017Interaction}. This yields:
\[U\left( h \right) = \pi \varepsilon {\rho _{hBN}}{\rho _C}\left( {\frac{{4{\sigma ^{12}}}}{{5{h^{10}}}} - \frac{{2{\sigma ^6}}}{{{h^4}}}} \right),\]
where $h$ is the interlayer distance, $\rho_{hBN}$ and $\rho_C$ are the two-dimensional atomic densities (number of atoms per area), in hBN and graphene respectively and $\varepsilon$ and $\sigma$ are the Lennard-Jones parameters. Since the hBN substrate consists of two different atoms we assume that $\sigma_{CB} = \sigma_{CN}$ and let $\varepsilon$ be the average Lennard-Jones binding energy: $\varepsilon  = \left( {{\varepsilon _{CB}} + {\varepsilon _{CN}}} \right)/2$. The equilibrium interlayer distance, where $U$ is minimal, is $h=\sigma$. The vdW energy per area after buckling is therefore: $U(y(l)+\sigma)$. The total vdW energy cost of buckling becomes: 
\begin{equation} \label{eq1}
\begin{split}
\Delta {E_{vdW}} &= \frac{w}{2}\int_0^\lambda  {\left[ {U\left( {y\left( l \right) + \sigma } \right) - U\left( \sigma  \right)} \right]dl} \\
 &= \frac{9}{2}\pi \varepsilon {\rho _{hBN}}{\rho _C}\lambda w{A^2}.
\end{split}
\end{equation}
Here we used a Taylor approximation to the lowest order in $A$, since we consider the onset of buckling where it is reasonable to assume that $A$ is small. This result is in agreement with Ref. \cite{Koskinen_2013}. 
The total energy change at the onset of buckling is thus:
\[\Delta E = {A^2}\frac{w}{2}\left( { - \frac{{k{\pi ^2}}}{{2\lambda }}\varepsilon  + \frac{{D{\pi ^4}}}{{{\lambda ^3}}} + 9\pi \varepsilon {\rho _{hBN}}{\rho _C}\lambda } \right).\]
Buckling occurs when $\Delta E\left( \varepsilon  \right) = 0$. The energy of the buckled GNR is minimized with respect to $\lambda$. This yields an expression for the critical strain $\varepsilon_b$ where the ribbon buckles:
\[{\varepsilon _b} = \frac{{12}}{k}\sqrt {\pi D{\rho _C}} \sqrt {\varepsilon {\rho _{hBN}}} .\]
According to the model, $\varepsilon_b = 2.5\%$ for graphene ribbons on hBN and $\varepsilon_b = 2.2\%$ for graphene ribbons on graphene. In the experiments in the maintext we find $\varepsilon_b \sim 3\%$, but it should be noted that the ribbons are probably pushed a small distance beyond the buckling point, due to the discrete steps used during nano-manipulation.

\newpage
\section{Additional data}

\begin{figure*}[h]
	\includegraphics[width=1.00\linewidth]{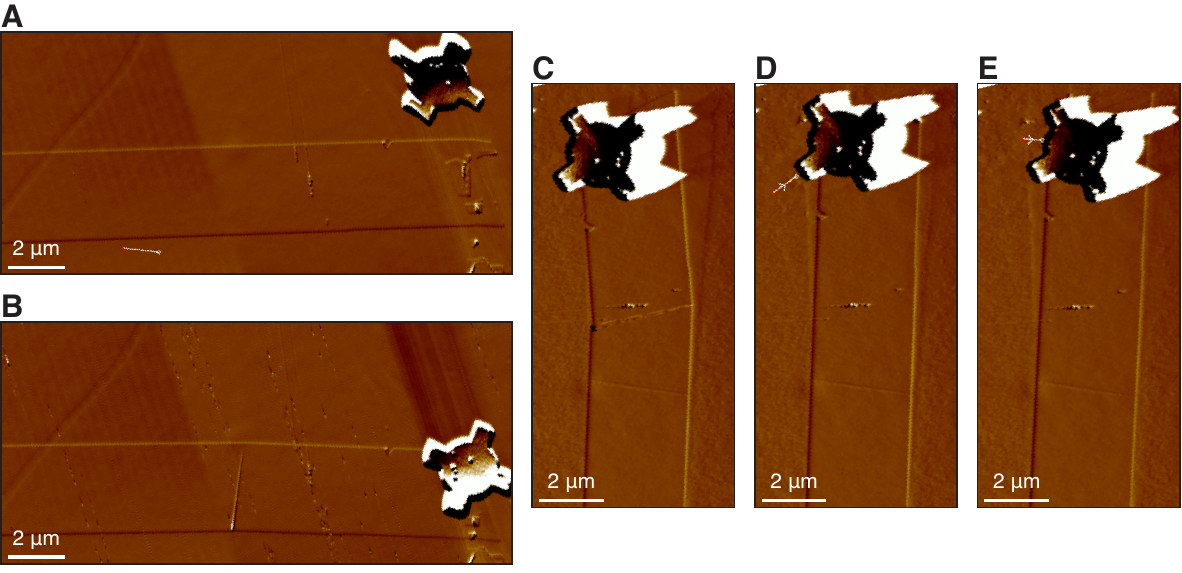}
	\caption[]{\textbf{Buckling and reversibility}
	A,B) The upper panels shows an AFM scan of a MLG before bending, while the lower panel shows the same ribbon after a displacement of leading to buckling,
	C) Buckled MLG after pushing,
 	D) Same MLG as in C), the buckle has been undone by pushing on the MLG with the manipulator as shown by the arrow on the top left,
 	E) MLG ribbon pushed in the other direction after un-buckling in D).}\label{fig:SI2}
\end{figure*}

\begin{figure*}[h]
	\includegraphics[width=1\linewidth]{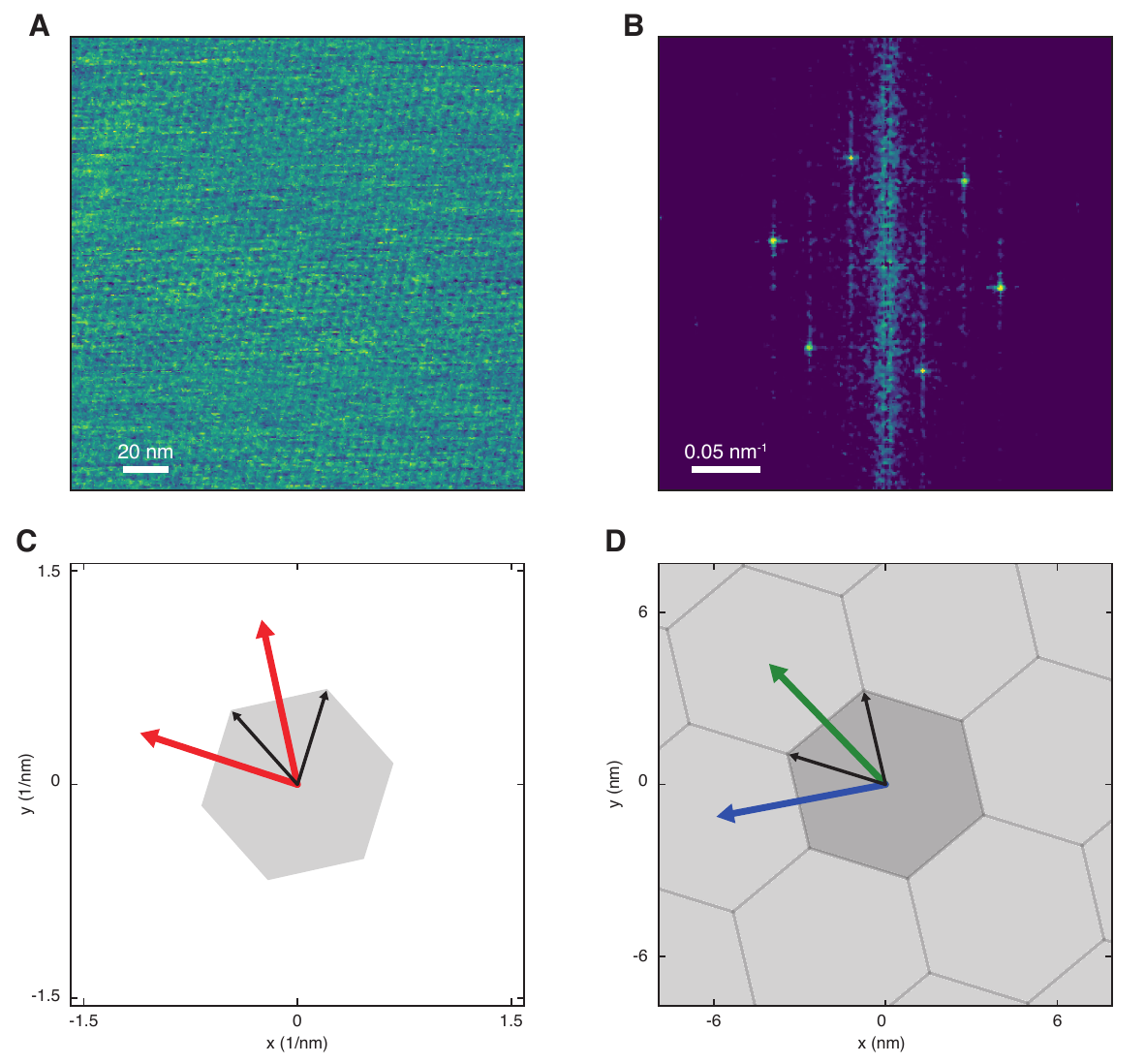}
	\caption[]{\textbf{Strain fitting}
	A) PFM scan of a bMLG on hBN,
	B) FFT of scan shown in A),
	C) Reciprocal space of the bMLG on hBN system shown in A). Red arrows show the fit of to the peak position in B) allowing us to extract the strain and twist angle, yielding a twist angle of 2.3$^{\circ}$ and a strain of 0.17\,\%
 	D) Real space calculated from the fit to the moiré vectors in C).}\label{fig:SI2}
\end{figure*}

\begin{figure*}[h]
	\includegraphics[width=1\linewidth]{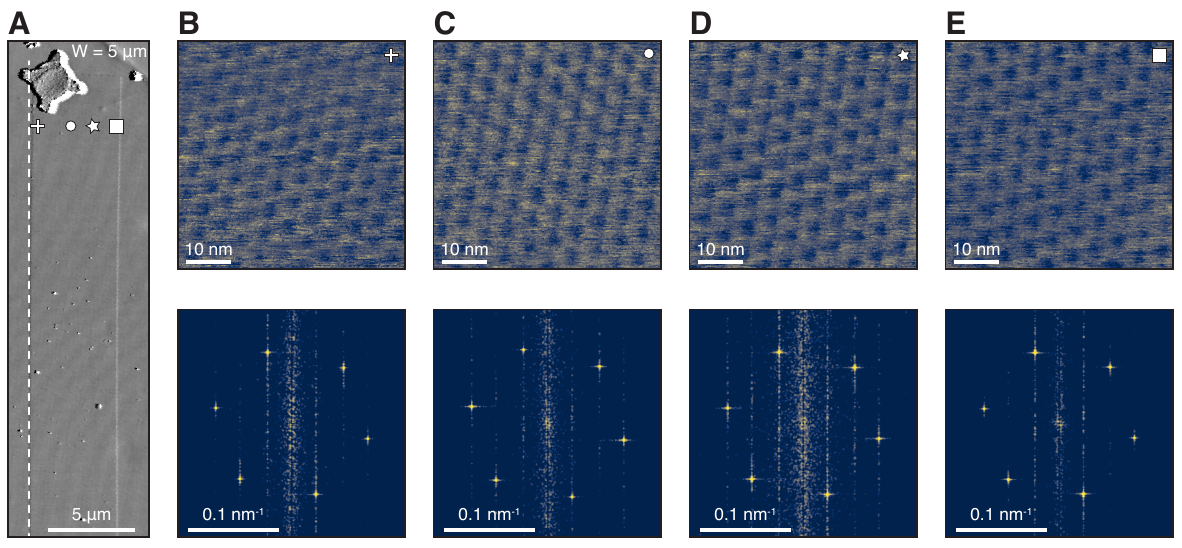}
	\caption[]{\textbf{PFM across a wide ribbon}
	A) AFM image of the bMGL,
	B,C,D,E) PFM scans and their corresponding FFT in the lower panel of positions at constant $x$ (or twist angle) and varying $y$ (or strain) showing very little strain evolution across the ribbon.}\label{fig:SI3}
\end{figure*}

\begin{figure*}[h]
	\includegraphics[width=1\linewidth]{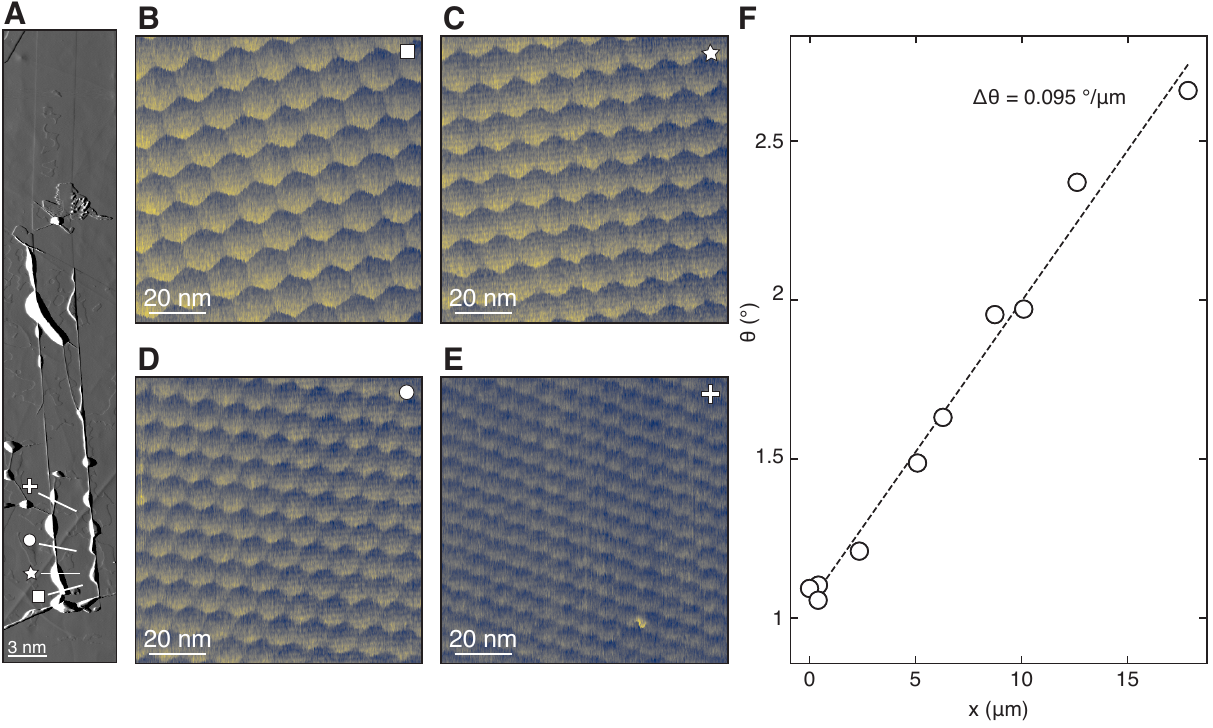}
	\caption[]{\textbf{Angle evolution of bBLG}
	A) AFM image of the bBLG device,
	B,C,D,E) LFM scans with positions marked in A) showing a twist angle evolution from lower twist angle in B) to higher twist angle in E),
	F) Summary of the twist angle evolution along the device shown in A) extracted from the moiré wavelength calculated using LFM scans. The global twist angle evolution is 0.095 . }\label{fig:SI6}
\end{figure*}

\begin{figure*}[h]
	\includegraphics[width = 1\textwidth]{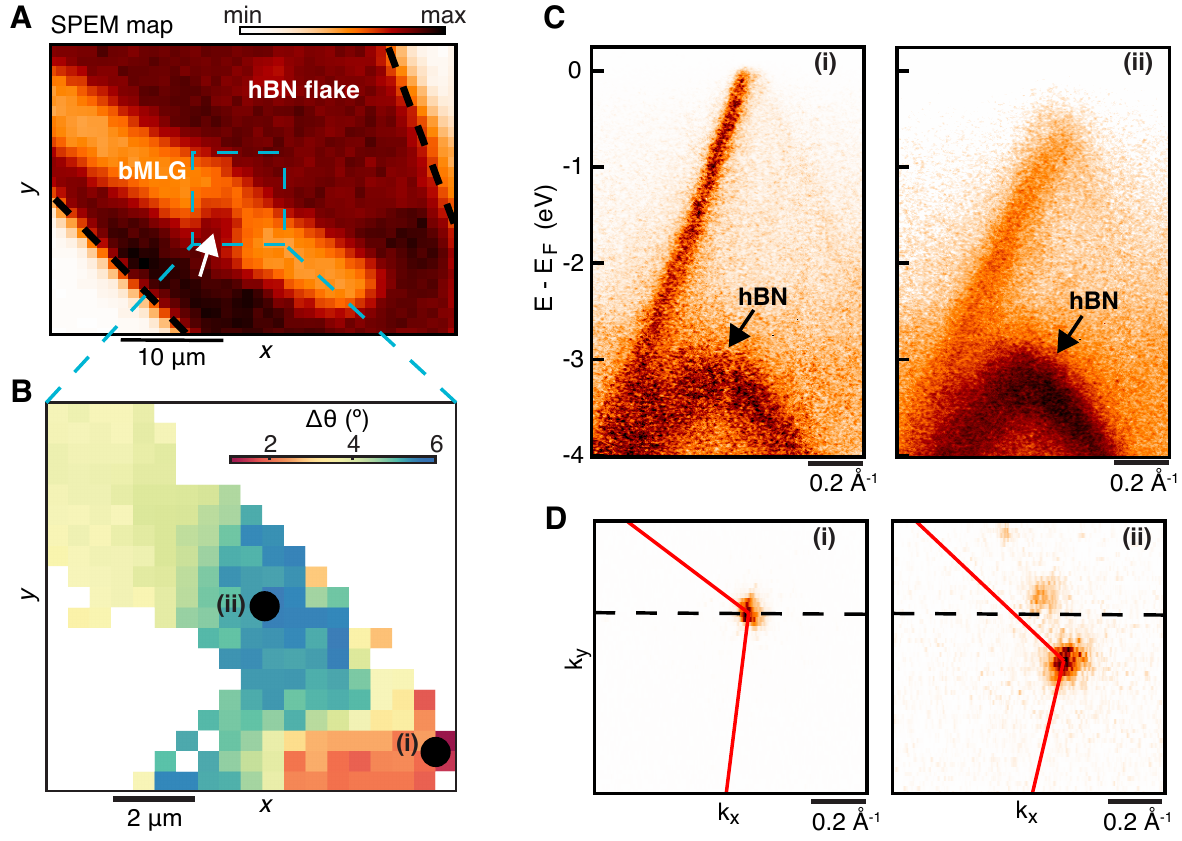}
	\caption[]{\textbf{NanoARPES of a 7\,$\mu$m wide bMLG on hBN}
	A) Scanning photoemission microscopy (SPEM) map of bMLG on hBN. The arrow marks the position of the nano-manipulator, which is located on top of the bMLG. The dashed black lines outline the hBN flake.
	B) Relative twist-angle mapping of the bMLG within the area marked by a blue dashed square in A). The shift in twist angle is determined via the displacement of the graphene Dirac cone in each position of the SPEM map.
	C) ARPES spectra of bMLG and hBN at different twist angles from the locations indicated in B). As expected for such large ribbon, the bandstructure of graphene remains unmodified by the bending as the strain remains low. Only the twist angle varies along the device.
	D) Constant energy cuts at the Fermi energy. The red lines demarcate the main Brillouin zone corners within the measurement window. The dashed horizontal line corresponds to the $E(k_x)$ cut measured in A)-C). The faint pockets of intensity in (ii) emerge from rotational domains that shattered from the main bMLG during bending.}\label{fig:SI_nanoarpes}
\end{figure*}

\begin{figure*}[h]
	\includegraphics[width=1\linewidth]{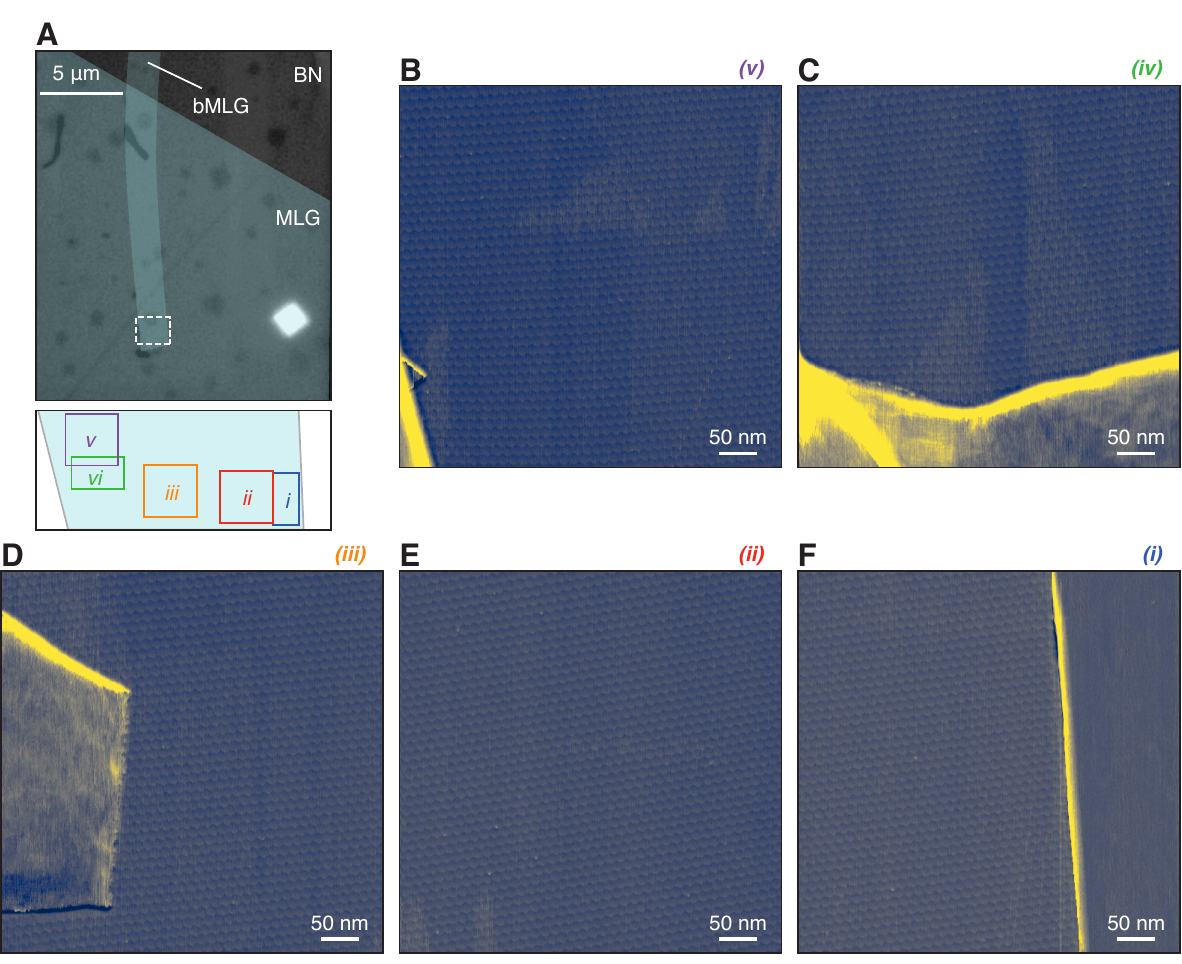}
	\caption[]{\textbf{LFM of bBLG around magic angle}
	A) Microscope image of the bBLG device with LFM scans positions shown in the lower panel,
	B,C,D,E,F) LFM scans around magic angle across the bBLG device shown in A).}\label{fig:SI7}
\end{figure*}

\begin{figure*}[h]
	\includegraphics[width=1\linewidth]{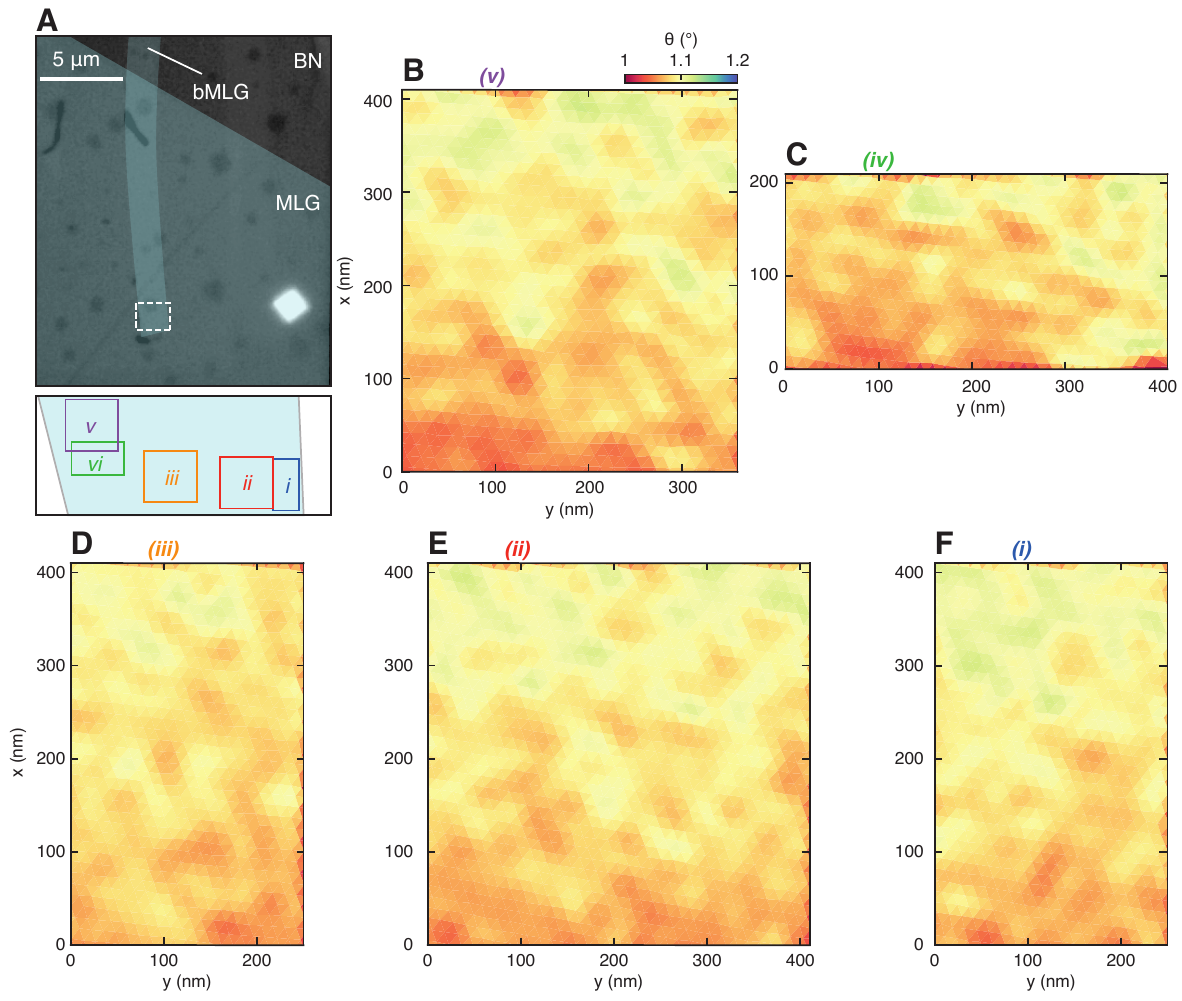}
	\caption[]{\textbf{Local twist-angle mapping}
	A) Microscope image of the bBLG device with LFM scans positions shown in the lower panel,
	B,C,D,E,F) Twist-angle mapping using the next-neighbor average distance to calculate the moiré wavelength for each LFM scans shown in A).}\label{fig:SI8}
\end{figure*}

\begin{figure*}[h]
	\includegraphics[width=1\linewidth]{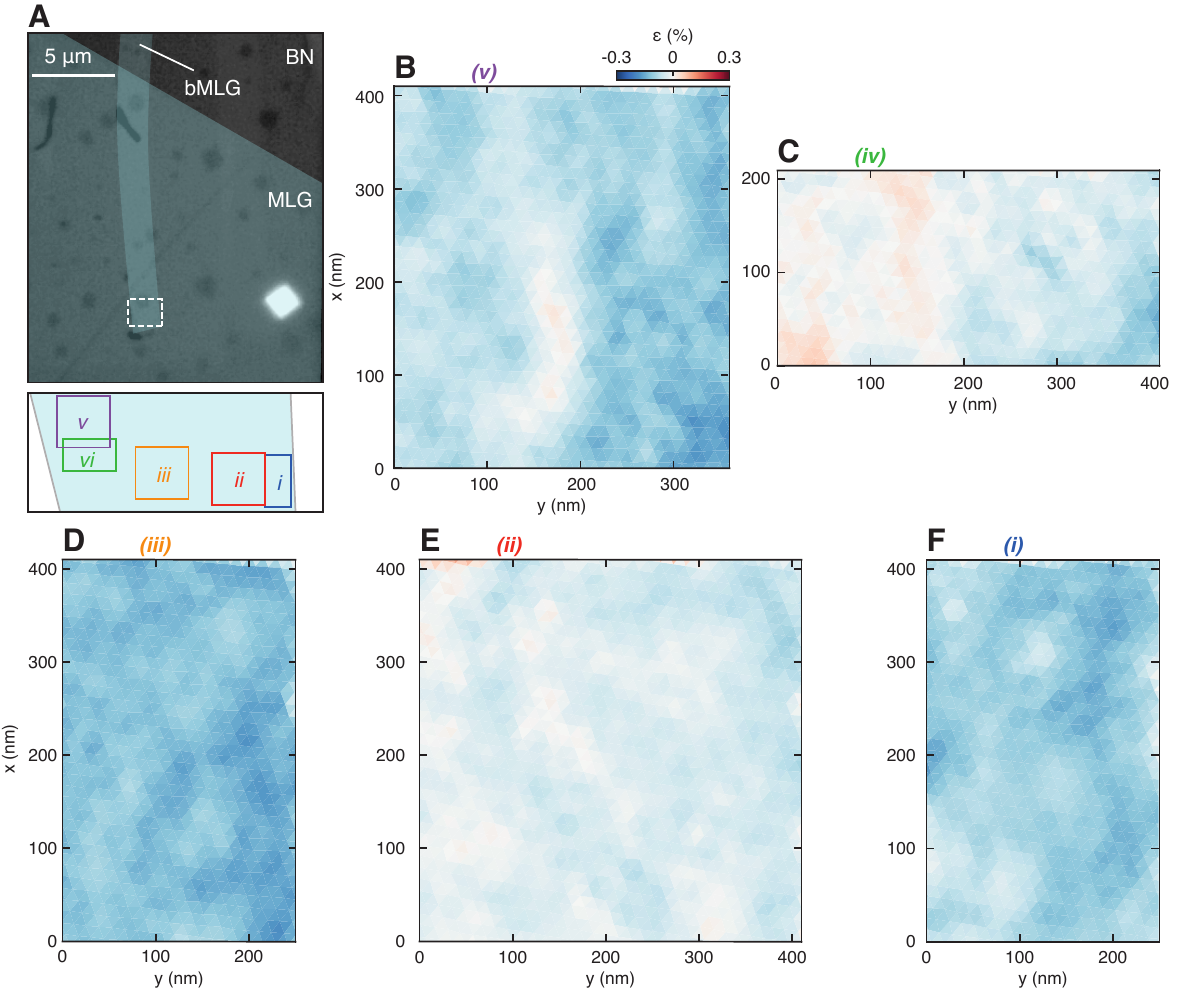}
	\caption[]{\textbf{Local strain mapping}
	A) Microscope image of the bBLG device with LFM scans positions shown in the lower panel,
	B,C,D,E,F) Strain mapping using the next-neighbor average distance to calculate the moiré wavelength for each LFM scans shown in A).}\label{fig:SI9}
\end{figure*}

\begin{figure*}[h]
	\includegraphics[width=1\linewidth]{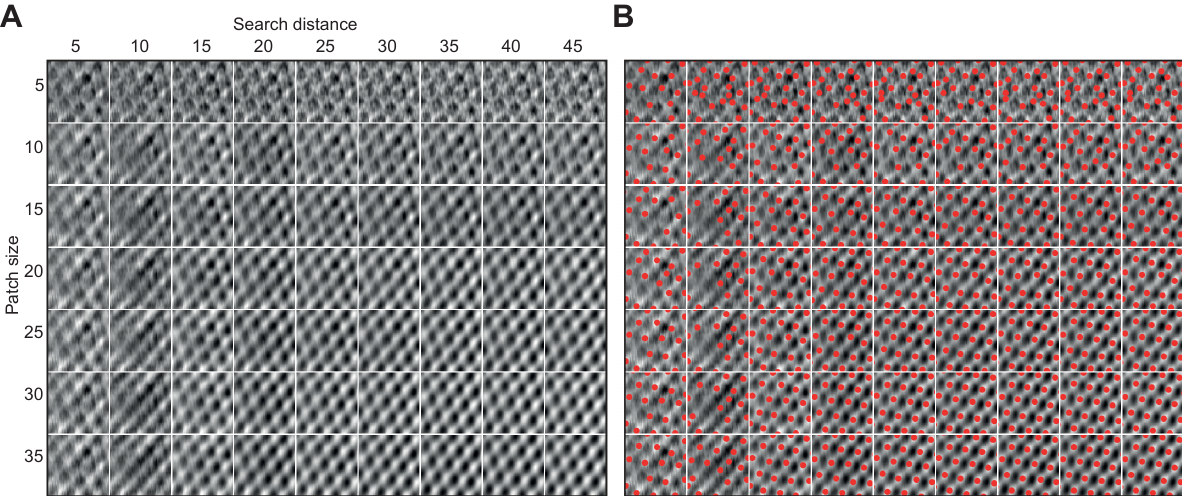}
	\caption[]{\textbf{Denoising}: Non-local means (NLM) denoising \cite{buades_non-local_2011} extends on the usual median filtering, by comparing the median pixel value of similar image patches within a given search distance, and thus works particularly well for denoising data with periodic patches. In panel A we compare the results of various patch sizes and search distances, and in panel B we show the resulting detected superlattice sites. For search distances and patch sizes above $\sim30$\,px we reliably identify all superlattice sites. Each window is $\sim50$ x $50\,$px$^2$ at $\sim0.977\,$nm/px.}\label{fig:SI10}
\end{figure*}

\clearpage
\bibliographystyle{naturemag_noURL}


\end{document}